\documentclass[conference]{IEEEtran}
\usepackage{cite}
\usepackage{amsmath,amssymb,amsfonts}
\usepackage{algorithmic}
\usepackage{graphicx}
\usepackage[table,xcdraw,HTML]{xcolor}
\usepackage{textcomp}
\def\BibTeX{{\rm B\kern-.05em{\sc i\kern-.025em b}\kern-.08em
    T\kern-.1667em\lower.7ex\hbox{E}\kern-.125emX}}
\usepackage[T1]{fontenc}
\usepackage{longtable}
\usepackage{array}
\usepackage{tabularx}
\usepackage{framed}
\usepackage[multiple]{footmisc} 
\usepackage{ragged2e}
\usepackage{xurl}
\usepackage{hanging}
\usepackage{comment}
\usepackage{multirow}
\usepackage{booktabs}
\usepackage{tabularx}
\usepackage[most]{tcolorbox}
\usepackage{float}
\usepackage{pdfpages}
\newcommand{\cmmnt}[1]{\ignorespaces}
\usepackage{hyperref}

\definecolor{lightgreen}{HTML}{E6F4EA}
\definecolor{bordercolor}{HTML}{006400}

\begin{document}

\title{Sustainability Flags for the Identification of Sustainability Posts in Q\&A Platforms\\
}
\author{\IEEEauthorblockN{1\textsuperscript{st} Sahar Ahmadisakha}
\IEEEauthorblockA{\textit{Faculty of Science and Engineering} \\
\textit{University of Groningen}\\
Groningen, Netherlands \\
0000-0002-1485-4512}
\and
\IEEEauthorblockN{2\textsuperscript{nd} Lech Bialek}
\IEEEauthorblockA{\textit{Faculty of Science and Engineering} \\
\textit{University of Groningen}\\
Groningen, Netherlands \\
0000-0002-6856-1862}
\and
\IEEEauthorblockN{3\textsuperscript{rd} Mohamed Soliman}
\IEEEauthorblockA{\textit{Heinz Nixdorf Institut} \\
\textit{Paderborn University}\\
Paderborn, Germany \\
0000-0002-6638-3732}
\and
\IEEEauthorblockN{4\textsuperscript{th} Vasilios Andrikopoulos}
\IEEEauthorblockA{\textit{Faculty of Science and Engineering} \\
\textit{University of Groningen}\\
Groningen, Netherlands \\
0000-0001-7937-0247}
}

\maketitle
\begin{abstract}
In recent years, sustainability in software systems has gained significant attention, especially with the rise of cloud computing and the shift towards cloud-based architectures. This shift has intensified the need to identify sustainability in architectural discussions to take informed architectural decisions. One source to see these decisions is in online Q\&A forums among practitioners' discussions. However, recognizing sustainability concepts within software practitioners' discussions remains challenging due to the lack of clear and distinct guidelines for this task. To address this issue, we introduce the notion of sustainability flags as pointers in relevant discussions, developed through thematic analysis of multiple sustainability best practices from cloud providers. This study further evaluates the effectiveness of these flags in identifying sustainability within cloud architecture posts, using a controlled experiment. Preliminary results suggest that the use of flags results in classifying fewer posts as sustainability-related compared to a control group, with moderately higher certainty and significantly improved performance. Moreover, sustainability flags are perceived as more useful and understandable than relying solely on definitions for identifying sustainability.
\end{abstract}
\begin{IEEEkeywords}
design science, empirical software engineering, experiment, sustainability, cloud computing, Stack Exchange.
\end{IEEEkeywords}

\section{Introduction}
Software engineering plays a crucial role in ensuring the sustainability of software systems and includes four dimensions that defines in Lago et. al~\cite{lago-2015}: \textit{technical} dimension, referring to a software system's ability to evolve and remain in use over the long term; \textit{economic} dimension, concerning the preservation and creation of capital and value; \textit{social} dimension, focusing on the continuity of communities using the software system; and \textit{environmental} dimension, aiming to minimize the system's impact on natural resources.

The growing popularity of cloud computing~\cite{ramchand2021enterprise, golightly2022adoption}, coupled with the increasing importance of sustainability~\cite{venters2023sustainable}, has further underscored the need to better understand the intersection of these two domains~\cite{andrikopoulos2021}. Following the introduction of cloud computing in 2006 and the development of guidelines for building software systems on top of it~\cite{cito2015making} using large-scale data centers, the discussion around sustainability in cloud software systems began, particularly from an environmental perspective~\cite{ferrer2012optimis, garg2013framework, wajid2015achieving}. Over time, the integration of sustainability concerns into \textit{cloud architecture, that is of architecting software to run in the cloud}, has gained increasing attention~\cite{andrikopoulos2021}, as explored in studies such as \cite{vos2022architectural, procaccianti-2015, ahmadisakha2024architecting, andrikopoulos2021, ahmadisakha2024mining} which are beyond focusing only on the environmental dimension.

As the intersection between cloud architecture and sustainability gains attention, practitioners as well need to be aware of sustainability in their decision-making discussions~\cite{andrikopoulos2021} so they can make informed architectural decisions with regard to sustainability in their daily practice~\cite{ahmadisakha2023pragmatic}. This makes identifying sustainability in the architectural discussions crucial. To start solving this problem, we begin with identifying sustainability in the discussions (posts) of online Q\&A platforms to further help practitioners identify sustainability in their daily architectural decisions. We choose online Q\&A platforms to start with since they facilitate informed architectural decisions through community discussions~\cite{soliman2016architectural} and they have been examined in multiple research projects for exploring architectural decisions~\cite{de2023characterizing, bi2021mining, soliman2016architectural}. They are also a valuable source of exchanging idea, sharing decisions, and are known to be the most valuable sources of information and insight~\cite{de2022developers}.

A key challenge in this process is identifying whether a post in a Q\&A forum is actually related to sustainability. Despite the growing interest in sustainability, there is still no clear method or guideline for recognizing sustainability-related content in any of the source of textual content (including posts from Q\&A forums). While definitions of sustainability dimensions (as presented in the initial paragraph of the same section) provide a valuable starting point for this task, they are not sufficient for identification~\cite{ahmadisakha2024mining}. This is because definitions, by nature, are open to interpretation and often require a level of expertise~\cite{ahmadisakha2023pragmatic, ahmadisakha2024architecting}. As a result, it remains difficult to identify sustainability-related posts. 

Given the deficiencies of only relying on sustainability definitions and in light of the lack of a method or guideline in the literature to help us through sustainability identification, our \textbf{goal} is to help identify sustainability within the cloud architectural posts in Q\&A forums to further help practitioners identify sustainability in their decisions. This goal leads us to first \textit{introduce the notion of \textbf{sustainability flags}}, that is, criteria designed to flag posts as sustainability-related. Second,\textit{ we want to see how effective this notion is in the sustainability identification task.}

These flags are derived from the \textit{sustainability best practices (BP)} of hyperscalers (cloud service providers) such as Amazon Web Services (AWS) and Microsoft Azure (MA) in their well-architected frameworks (WAF). These frameworks are knowledge bases that describe key concepts, design principles, and architectural BPs for designing and running workloads on top of the hyperscalers. Both frameworks contain sustainability pillars,~\cite{waf-aws} and~\cite{waf-az}, that will be used as the two main sources for extracting sustainability flags in this work. We select these sources for extracting the flags as they align with our goal of identifying sustainability in cloud architectural posts and are highly relevant to the cloud architectural context. A post is a cloud architectural post if the architectural concern or solution discussed in it is about or applicable to applications based on the cloud computing environment. We extract these flags through thematic analysis of sustainability BPs and test their effectiveness in a controlled experiment. Through a comparison between a test and a control group, we assess whether sustainability flags offer an advantage in identifying sustainability-related content over using only the sustainability dimensions' definitions.

The main \textbf{contribution} of this work is therefore the introduction and evaluation of sustainability flags, which aid in classifying and analyzing sustainability-related posts in Q\&A forums. The \textbf{scope} of the study, as well as the proposed flags, is limited to cloud architectural posts, as the flags originate from cloud architectural BPs and will be applied exclusively to a dataset containing only cloud architectural posts.
This paper is structured as follows: In Section~\ref{sec:rqs}, we outline the research questions and methods. Section~\ref{sec:thematic} presents the design of the thematic analysis and Section~\ref{sec:flags} its results. Sections~\ref{sec:experiment}, ~\ref{sec:experiment-exec}, and ~\ref{sec:experiment-analysis} detail the experiment's design, execution, and results. In Section~\ref{sec:discussion}, we present discussion points. Section~\ref{sec:ttv} addresses threats to validity. Related works are discussed in Section~\ref{sec:related}, and Section~\ref{sec:conclusion} concludes the work.
%
\begin{figure*}[t]
\centering
\includegraphics[width=1.7\columnwidth]{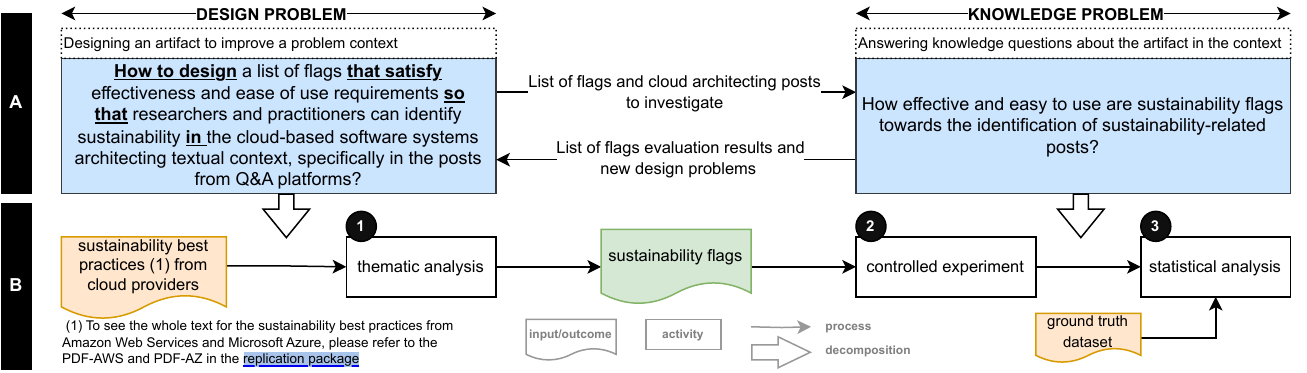}
\caption{The study design process in the design science framework. Row \textbf{A} shows the overall connections of the research questions and row \textbf{B} shows the detailed activities done during each one.}
\label{fig:design-sci}
\end{figure*}
\section{Study Overall Design}
\label{sec:rqs}
\subsection{Research Goals and Questions}
The goal of this work is to help identify sustainability within architecturally relevant posts about cloud computing, such as those found in posts on Q\&A platforms like Stack Exchange. We believe this goal can be better explained using the design science framework stated by Wieringa~\cite{wieringa2014design}. According to this framework, we break the goal into two different questions, one design problem (or technical research question) and a knowledge question. The former calls for a change in the problem context by creating an artifact or a solution, while the latter asks for knowledge about the artifact in the context. 
\noindent Following this framework, the first research question (RQ1), is stated as a \textit{design problem} and the second research question (RQ2) is presented as a \textit{knowledge question}, as summarized in the ``A'' row in Fig.~\ref{fig:design-sci}. 
\begin{tcolorbox}[
    colback=lightgreen,  
    colframe=bordercolor, 
    width=\columnwidth, 
    sharp corners, 
    boxrule=0.6mm, 
    left=3pt, right=3pt, top=3pt, bottom=3pt, 
    before skip=5pt, after skip=5pt, 
    breakable
]\textit{\textbf{RQ1:} \textbf{How to design} a list of flags \textbf{that satisfy} effectiveness and ease of use requirements \textbf{so that} researchers and practitioners can identify sustainability \textbf{in} the cloud-based software systems architecting textual context, specifically in the posts from Q\&A platforms?}
\end{tcolorbox}
Identification of sustainability by solely relying on the general definition might be misleading. We believe that we need a more tangible and definite way to identify it in the content of the posts. \emph{Sustainability flags} as a list of criteria serve as guidance to help identify sustainability within posts. These flags are to be derived from established BPs for sustainability introduced by major hyperscalers like AWS and MA. These BPs try to solve the problems that lead to sustainability impacts. We think that if we focus on identifying factors within the BPs that are affected or altered to improve sustainability, we actually identify sustainability flags. 
The next question asks how successful is this notion:

\begin{tcolorbox}[
    colback=lightgreen,  
    colframe=bordercolor, 
    width=\columnwidth, 
    sharp corners, 
    boxrule=0.6mm, 
    left=3pt, right=3pt, top=3pt, bottom=3pt, 
    before skip=5pt, after skip=5pt, 
    breakable
]\textit{\textbf{RQ2:} How effective and easy to use are sustainability flags for the identification of sustainability-related posts?}\end{tcolorbox}

We cannot assert that simply because we derive certain concepts (flags) from relevant context, they are sufficiently effective for identifying sustainability or that they are easy to use. To evaluate their effectiveness and ease of use, the adoption of the flags needs to be tested and measured using appropriate metrics such as certainty in their use. The goal of this RQ is to evaluate whether the sustainability flags we developed for identifying sustainability in the previous RQ are effective and easy to use. 
In the following, we outline which methods we use to answer these questions.
\subsection{Methods used}
\label{sec:methods}
This study employs two research methods--namely thematic analysis (for RQ1) and controlled experiment (for RQ2)--and utilizes both qualitative and quantitative approaches to analyze the data. The details for each are presented in row ``B'', Fig.~\ref{fig:design-sci}.

We choose thematic analysis~\cite{braun2012thematic} to address RQ1 and extract the sustainability flags because it is well-suited for handling context-dependent data~\cite{vaismoradi2013content},~\cite{fereday2006demonstrating}. Our data, being specific to cloud-based applications, benefits from this approach. Additionally, this method has been successfully applied in related fields, such as energy efficiency and software applications in~\cite{cruz2019catalog}.
The established sustainability BPs introduced by major cloud providers like AWS and MA are to be used as input to this process.

Once we finish extracting the necessary flags, we evaluate them to determine their effectiveness in the spirit of RQ2. The method used for conducting the evaluation is a controlled experiment~\cite{easterbrook2008selecting} together with statistical tests as in~\cite{kitchenham2017robust}. We use the guidelines for conducting a controlled experiment presented in~\cite{kitchenham2002preliminary, wohlin2012experimentation}. We decide to present the design, execution, and results of each method per RQ in separate sections. In Sections~\ref{sec:thematic} and~\ref{sec:experiment} we discuss the process for RQ1 and RQ2, respectively.

\section{Thematic Analysis: Towards Extracting Sustainability Flags}
\label{sec:thematic}
To conduct our analysis, we closely followed the guidelines outlined by~\cite{cruzes2011} and \textit{reported} each phase as defined by~\cite{braun2012thematic}. Besides, we employ the \textit{coding} concept as stated by~\cite{williams2019art} alongside the thematic analysis to build upon our themes. The whole process is collaboratively done by three researchers, using a qualitative analysis tool, Atlas.ti~\cite{atlasti}. A protocol was established and shared with all researchers before beginning each phase to ensure rigorous adherence to the guidelines. As per~\cite{braun2012thematic}, the protocol consists of the following \textit{phases}:
\subsubsection{Familiarization With the Data} In this phase, three researchers review all 92 best practices (28 from AWS and 66 from MA) to familiarize themselves with the data, consulting additional readings linked within the best practices context to clarify any ambiguities.
\subsubsection{Generating Initial Codes} In this phase, two researchers select 15 BPs (6 from AWS and 9 from MA) to extract codes in parallel. They then discuss these codes to align their understanding of the analysis goal, minimize subjectivity, and identify differences; \textit{no significant discrepancies are noted. Cohen’s Kappa coefficient, a standard measure for inter-rater agreement \cite{cohen1960coefficient} is calculated as: \textbf{0.7059}.\footnote{According to established guidelines by Landis and Koch \cite{landis1977measurement}, Kappa value can be evaluated as: (0.00-0.20: slight agreement), (0.21-0.40: fair agreement), (0.41-0.60: moderate agreement), (0.61-0.80: substantial agreement), (0.81-0.99: almost perfect agreement).} This value is calculated per code that is assigned per BP.} Following this, each researcher independently extracts codes from all BPs (open coding), resulting in 112 codes from one researcher and 99 from the other. These codes are then collaboratively and iteratively consolidated into 63, after combining duplicates and merging those representing similar concepts (axial coding).
\subsubsection{Searching for Themes} In this phase, with only one researcher involved, themes, are identified and organized into an initial set. The researcher reviews the codes, grouping them into meaningful subsets and assigning a theme to each subset (selective coding). Themes are crafted to be clear, distinct, and representative of the codes, which ensures they are understandable by a non-specialist audience in sustainability.
\subsubsection{Reviewing the Themes} In this phase, two researchers collaborate to discuss and refine the flags. They review the themes alongside the original codes and the axial codes once more to determine if any adjustments are needed. During this process, two themes are merged with others to ensure the themes are distinct and convey clear concepts without overlap.
\subsubsection{Defining and Naming Themes (the flags)} In this phase, one researcher begins by naming and defining the flags based on the identified themes. A second researcher then reviews the names and definitions to minimize subjectivity. Together, they discuss potential overlaps among flags and refine the definition structure. This process results in a final set of \textbf{seven flags}.
\subsubsection{Producing the Report} 
\begin{figure}[t]
\centering
\includegraphics[width=0.9\columnwidth]{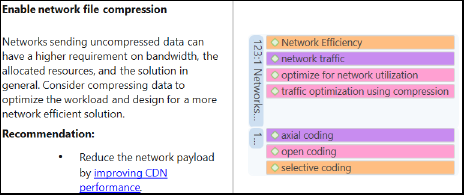}
\caption{An example of a BP together with the actual codes assigned to it.}
\label{fig:sample-bp}
\end{figure}

In this phase, three researchers extensively discuss the naming, definitions, and examples of each emerging theme. Through this process, we refine the definitions of two flags and adjust examples for two others to enhance clarity.
In all the above phases that more than one researcher is involved, although only minor discrepancies were observed but the \textbf{full agreement} is always met, which the details is presented in Sec.~\ref{sec:ttv}. In Fig.~\ref{fig:sample-bp}, we show an example of a BP, in this case one of the MA's BP. The example is an exact picture of existing analysis and coding done in Atlas.ti tool. Each level of the codes is highlighted with different colors and each BP has all three levels of open, axial, and selective coding. More details can be found in Atlas.ti's reports in the replication package, Sec.~\ref{sec:ttv}.\\
\vspace{-10pt}
\section{Sustainability Flags}
\label{sec:flags}
\noindent \textbf{Seven} extracted themes (sustainability flags) are reported here. Examples with links and citations are from the actual BPs.
\begin{tcolorbox}[
    colback=lightgreen,  
    colframe=bordercolor, 
    width=\columnwidth, 
    sharp corners, 
    boxrule=0.6mm, 
    left=5pt, right=5pt, top=5pt, bottom=5pt, 
    before skip=5pt, after skip=5pt, 
    breakable,
    fontupper={\fontsize{9}{9}\selectfont}
]
\noindent\textbf{Resource Efficiency:} Optimal use of computational and system resources (for instance: CPU, memory, and GPU) to ensure that applications and services run with minimal waste. This includes reducing idle resources, minimizing unnecessary computation, and rightsizing resources to match the actual workload. \textit{Example BPs:} \href{https://docs.aws.amazon.com/wellarchitected/latest/sustainability-pillar/sus_sus_software_a3.html}{Remove or refactor workload components with low or no use} like a drop in serverless function run time~\cite{aws-sw_arch}; \href{https://docs.aws.amazon.com/wellarchitected/latest/sustainability-pillar/sus_sus_software_a2.html}{Optimize software and architecture for asynchronous and scheduled jobs}~\cite{aws-sw_arch} like using queues to spread out workloads over the available resources in time and if needed in batches.
\vspace{3pt}
\hrule width 0.95\columnwidth \vspace{3pt}

\noindent\textbf{Network Efficiency:} The optimization of network usage by minimizing unnecessary data transmission, reducing bandwidth overhead, and optimizing routing paths to create communication with minimal energy consumption. \textit{Example BPs:} \href{https://learn.microsoft.com/en-us/azure/well-architected/sustainability/sustainability-application-design#evaluate-server-side-vs-client-side-rendering}{Evaluate server-side vs. client-side rendering}~\cite{azure-app_design}; \href{https://learn.microsoft.com/en-us/azure/well-architected/sustainability/sustainability-networking#make-use-of-a-cdn}{Making use of a CDN (Content Delivery Network)}~\cite{azure-network}.
\hrule width 0.95\columnwidth \vspace{3pt}

\noindent\textbf{Storage Efficiency:} The effective use and management of storage resources by reducing data duplication, compressing stored data, and eliminating unnecessary or redundant storage usage. This includes efficient data retention policies and optimizing storage access patterns. \textit{Example BPs:} \href{https://learn.microsoft.com/en-us/azure/well-architected/sustainability/sustainability-storage#enable-storage-compression}{Enable storage compression}~\cite{azure-storage}; \href{https://docs.aws.amazon.com/wellarchitected/latest/sustainability-pillar/sus_sus_data_a3.html}{Use technologies that support data access and storage patterns}~\cite{aws-data_mgmt} like hot and cold data tiers.
\vspace{3pt}
\hrule width 0.95\columnwidth \vspace{3pt}

\noindent\textbf{Code Efficiency:} The optimization of software code to reduce unnecessary processing, improve execution speed, and lower resource consumption (for example: CPU, memory). This optimization can be done through profiling, code review, and choice of the algorithm. Efficient code can minimize the system's workload, leading to reduced energy usage and improved overall performance. \textit{Example BPs:} \href{https://learn.microsoft.com/en-us/azure/well-architected/sustainability/sustainability-application-design#improve-api-efficiency}{Improve API efficiency}~\cite{azure-app_design} like Request throttling techniques; \href{https://learn.microsoft.com/en-us/azure/well-architected/sustainability/sustainability-application-design#leverage-cloud-native-design-patterns}{Leverage cloud native design patterns}~\cite{azure-app_design, aws-pattern} like \href{https://learn.microsoft.com/en-us/azure/architecture/patterns/ambassador}{Ambassador}~\cite{azure-pattern}.
\vspace{3pt}
\hrule width 0.95\columnwidth \vspace{3pt}

\noindent\textbf{Infrastructure Efficiency:} The efficient use of cloud and physical infrastructure, which ensures that computing resources (such as virtual machines, containers, and physical servers) are optimally utilized without over-provisioning or under-utilization. It includes selecting the appropriate infrastructure components for specific workloads. \textit{Example BPs:} \href{https://learn.microsoft.com/en-us/azure/well-architected/sustainability/sustainability-application-platform#containerize-workloads-where-applicable}{Containerize workloads where applicable}~\cite{azure-app_platform}; \href{https://docs.aws.amazon.com/wellarchitected/latest/sustainability-pillar/sus_sus_hardware_a2.html}{Using tailored instance types, example: fault-tolerant ones}~\cite{aws-instances}, \href{https://docs.aws.amazon.com/wellarchitected/latest/sustainability-pillar/sus_sus_hardware_a4.html}{Using managed services}~\cite{azure-app_platform} to reduce infrastructure overhead since they are more optimized, and \href{https://learn.microsoft.com/en-us/azure/well-architected/sustainability/sustainability-application-platform#choose-data-centers-close-to-the-customer}{Deploying applications in regions closer to users}~\cite{azure-app_platform}.
\vspace{3pt}
\hrule width 0.95\columnwidth \vspace{3pt}

\noindent\textbf{Energy Efficiency:} The reduction of energy consumption across all layers of the computing stack by using energy-efficient hardware, optimizing system usage, and reducing overall power consumption through intelligent deployment strategies and workload management. This mostly includes using renewable energy sources where and when possible. \textit{Example BPs:} \href{https://learn.microsoft.com/en-us/azure/well-architected/sustainability/sustainability-application-platform#deploy-to-low-carbon-regions}{Deploying to low-carbon regions}~\cite{azure-app_platform} and~\cite{aws-region}; \href{https://learn.microsoft.com/en-us/azure/well-architected/sustainability/sustainability-application-platform#process-when-the-carbon-intensity-is-low}{Process when the carbon intensity is low}~\cite{azure-app_platform} like running the workload at night.
\vspace{3pt}
\hrule width 0.95\columnwidth \vspace{3pt}

\noindent\textbf{Dynamic Resource Allocation:} The automatic and real-time allocation or deallocation of system resources (for instance: compute, memory, storage) based on current demand. It helps to resources be available when needed but do not remain idle when demand is low. This reduces over-provisioning and prevents resource waste. This is also about resource usage, but primarily about allocation. \textit{Example BPs:} \href{https://learn.microsoft.com/en-us/azure/well-architected/sustainability/sustainability-application-platform#utilize-auto-scaling-and-bursting-capabilities}{Utilize auto-scaling and bursting capabilities}~\cite{azure-app_platform} or \href{https://docs.aws.amazon.com/wellarchitected/latest/sustainability-pillar/sus_sus_user_a2.html}{Scale workload infrastructure dynamically}~\cite{aws-demand}, either vertically or horizontally to dynamically adjust resource allocations based on workload intensity.
\end{tcolorbox}
For the interested reader, further details are provided in the replication package of this study, illustrating how the flags are interrelated, the strength of these relationships, and identifying which flags emerged most frequently.
%
\section{Flags Evaluation: Design of the Experiment}
\label{sec:experiment}
%
As per Section~\ref{sec:methods}, this experiment follows instructions from~\cite{kitchenham2002preliminary} and~\cite{wohlin2012experimentation}. The \textit{report of the experiment} is structured using the guidelines from Jedlitschka and Pfahl~\cite{jedlitschka2005reporting} and~\cite{shull2008guide}, that leads to some extent of redundancies. The \textit{Inferences, Lessons Learned}, and \textit{Interpretation} from these guidelines are in Section~\ref{sec:discussion}; \textit{Limitations} in Section~\ref{sec:ttv}, and \textit{Relation to Existing Evidence} in Section~\ref{sec:related}.
\begin{table*}[t]
\centering
\caption{Independent and Dependent Variables of our experiment}
\label{tab:variables}
\fontsize{6}{7}\selectfont
\begin{tabular}{p{6cm} p{1.5cm} p{1cm} p{8cm}}  
\toprule
\multicolumn{4}{c}{\textbf{Dependent variables}} \\
\midrule
\textbf{Description} & \textbf{Scale type} & \textbf{Unit} & \textbf{Range} \\
\midrule
Certainty of sustainability identification & Interval & - & Numerical Slider, 0 and 5 for the least/most certainty\\
Number of posts with sustainability identified & Ratio & Posts & Positive natural number\\
Performance of sustainability identification & Ratio & Proportion & Number between 0 and 1\\
Ease of use the sustainability flags & Ordinal & - & Five-point Likert Scale [Not at all to Extremely for usefulness and understandability] and [Very hard to Very easy for difficulty]\\
\midrule
\multicolumn{4}{c}{\textbf{Independent variables}} \\
\midrule
Group & Nominal & -. & Possible values: Flag Group, Base Group\\
Master track name \textit{(Track)} & Nominal & - & Possible values: Track X, Track Y, Track Z (anonymized)\\ 
Year of the master \textit{(Year)} & Nominal & - & 2 Possible values: First, Second \\
Sustainability experience \textit{(Experience)} & Nominal & - & 3  Possible values: Slides, Course, Prior (multiple choices are possible)\\
\bottomrule
\end{tabular}
\end{table*}
\subsection{Goal, hypotheses, parameters, and variables}
The objective of this \textbf{experiment} is made more specific using the Goal-Question-Metric~\cite{van2002goal} formulation:
\textit{\textbf{Analyze} the use of sustainability flags \textbf{for the purpose of} understanding their \underline{effectiveness} and \underline{ease of use}, \textbf{with respect to} identifying sustainability-related posts \textbf{from the point of view} of non-experts on sustainability \textbf{in the context of} cloud-based software systems architecting.}

In general, the goal of this experiment is to determine whether the identification of sustainability-related posts using sustainability flags results in a more effective and easier method compared to the method that relies solely on definitions and prior knowledge about sustainability. More specifically, the \textbf{effectiveness} is checked using the \textit{certainty} of the labeling by non-experts on whether a post is sustainability-related and the \textit{quantity (counts)} of the labeled posts as ``Yes'', together with the evaluation of the labeling \textit{performance} in terms of accuracy, precision, recall, and F1-score against a ground truth\footnote{A ground truth dataset, is the same dataset used in the Object section, which its posts are labeled as whether are sustainability relevant or not. The Subjects (students) have zero visibility of it. The process of labeling the ground truth is detailed in Sec.~\ref{sec:ttvInternal}} labeling of the posts. Lastly, \textbf{ease of use} is to be evaluated by metrics: \textit{usefulness, understandability,} and \textit{difficulty} of using the given treatment.
\subsubsection{Dependent Variables}
For our experiment and based on the goal and hypotheses that we formulated, we have four dependent variables (summarized in Table~\ref{tab:variables}):

The \textit{certainty} of identifying sustainability-related posts is measured through two questions posed for each post. We first ask: `\textit{`Do you think this post is sustainability-related?'}' and then for the same post we ask: \textit{``How certain are you about your response to the above question?''} For the first question, the non-expert labelers only have a Yes/No option. For the second question labelers respond using a Slider on a continuous scale, where 0 represents \textit{the least certainty} and 5 represents \textit{the most certainty}. The labelers are instructed that a higher number indicates a stronger certainty level. Each post receives this value from two different labelers.

To measure the \textit{quantity} or the \textit{number of posts} identified as sustainability-relevant, we simply count the posts labeled as ``Yes'' by each group (test and control) without checking their correctness. To evaluate the \textit{performance} of both groups in identifying sustainability, we established a ground truth by first labeling the dataset of posts which is meant to be used as the object in this experiment. The dataset and ground truth will be further explained in Section~\ref{sec:ttvInternal}. We then compared each group’s results against this ground truth, using the metrics of accuracy, recall, precision, and F1-score~\cite{yacouby2020probabilistic}. 
Accuracy measures the proportion of correct predictions overall. Precision indicates the proportion of true positives among all positive predictions, while recall focuses on the proportion of true positives among actual positives. F1-score combines precision and recall to evaluate the balance between the two.

The \textit{ease of use} of the treatment for identifying sustainability evaluated using the results of a \textit{check questionnaire} presented to the labelers at the end of their assigned task. This questionnaire will assess aspects such as \textit{usefulness}, \textit{understandability}, and the \textit{difficulty} of using the treatments in identifying sustainability, shown in Table~\ref{tab:variables}. All of them are measured through a five-point Likert-scale range~\cite{joshi2015likert} which at least can be considered as an ordinal scale~\cite{sullivan2013analyzing}.
\subsubsection{Independent Variables}
We have a few independent variables presented in Table~\ref{tab:variables} related to the experience and characteristics of the experiment participants, which could potentially influence the outcomes of the experiment. 
The data includes the students' master track, year of study (1st or 2nd year), and their level of sustainability knowledge. Students may belong to one of three master tracks and gain sustainability knowledge through course slides, attending the sustainability lecture, or prior experience. They can select multiple options to describe their knowledge.
%
\subsubsection{Hypotheses}
\label{sec:hypotheses}
Now that we know all variables we can formulate our goal in terms of hypotheses, inspired by formulations in~\cite{pfleeger1995experimental, wettel2011software, bernardez2014controlled}:\\
\noindent \underline{$H_{01}$:} Using sustainability flags makes no difference in \textit{certainty} of identifying sustainability-related posts, compared to identification using only sustainability definition.\\
\underline{$H_{1}$:} The \textit{certainty} of identifying sustainability-related posts is higher when identification focuses on sustainability flags, compared to identification using only sustainability definition.\\
\underline{$H_{02}$:} Using sustainability flags makes no difference in the count of identified sustainability-related posts, compared to identification using only the sustainability definition.\\
\underline{$H_{2}$:} The count of identified sustainability-related posts is lower when the identification focuses on sustainability flags, compared to identification using only the sustainability definition.\\
\underline{$H_{03}$:} Identifying sustainability through the use of sustainability flags makes no difference in performance than relying on only the sustainability definition.\\
\underline{$H_{3}$:} Identifying sustainability through the use of sustainability flags yields better performance than relying on only the sustainability definition.
%

We do not formulate a hypothesis concerning the \textit{ease of use} since we are only interested in \textit{comparing} results between groups, without the need for showing statistical significance.
\subsection{Experiment design}
For the experiment design, we opt for a simple design as is recommended in~\cite{kitchenham2002preliminary}.
Among the four standard design types outlined in \cite{wohlin2012experimentation}, we select the simplest one: \textit{one factor with two treatments} and more specifically a \textit{completely randomized design}. In our case, the factor is the method of identifying sustainability-related posts, and the treatments are: (1) \emph{using sustainability flags} and (2) \emph{using sustainability definitions} to identify sustainability-related posts. \textbf{Both treatments are in the replication package, Sec.~\ref{sec:ttv}.}

In this design, each subject is exposed to one treatment. To achieve a balanced design, the best practice is to ideally have the same number of subjects per group. However, in our case the group size is 8 for the flag group and 9 for the base group, which was also utilized in previous works like~\cite{van2012supportive}. The number of posts given to the subjects in each treatment is near each other (28 to 31 posts per subject). Within both the flag and base groups, the sets of posts are divided entirely at random, and participants are randomly assigned to their respective treatments. This process is only carried out after applying proper blocking as is detailed in the following.
%
%
\subsubsection{Subjects}
\label{sec:subjects}
The experiment participants were \textit{Master's students} 
following a course on cloud computing taught at a computer science program at a European university.
This course spans cloud concepts from virtualization to architectural patterns and includes sustainability in cloud computing as one of the topics that it discusses. Students needed to attend the sustainability lecture that took place in the week leading to the experiment, and/or review the lecture slides to participate to the experiment. 
Of the 45 enrolled students, 20 attended the lecture, and 17 volunteered for the experiment. Participation was voluntary. There were no grading implications or incentives for joining.
\subsubsection{Object}
\label{sec:object}
The subjects were provided with a dataset of Stack Exchange posts related to cloud architecture discussions extracted from a work in the related literature~\cite{ahmadisakha2024mining}. 
The dataset consists of 192 posts from the Software Engineering forum of StackExchange, spanning 2011 to 2024. For our experiment, we focus on 121 posts from 2016 onward, aligning with the introduction of AWS' Well-Architected Framework in 2016. This filtering reinforces the link between the framework and the sustainability flags derived through thematic analysis. This dataset includes labels for each post indicating whether they are sustainability-related. These Yes/No labels, created through complete consensus among three researchers, are used for further analysis and are hidden from the subjects.
\subsubsection{Blocking and blinding}
Due to the moderately small sample size, students were assigned to the base and flag groups based on their master's program track, year of study, and sustainability experience rather than randomization~\cite{kitchenham2002preliminary}. This ensured balanced groups in terms of background and expertise, as shown in Table~\ref{tab:blocking}. In this experiment, \textit{materials are allocated blindly}. Participants are assigned to groups and receive materials via a pre-prepared questionnaire that ensures they remain unaware of their group, treatment, or the goals of other participants, though they know there are two groups. \textit{Blind marking} is managed by an additional researcher, with a third researcher renaming participant numbers to maintain anonymity. The response format does not reveal the treatment. However, \textit{blind analysis} is not feasible, as the experiment's researchers conduct the analysis, and group sizes would be disclosed regardless.
\subsubsection{Instrumentation}
The \textbf{demographic questionnaire}, used as a \textit{measurement instrument} in the \textit{pre-introduction step}, was completed by students to confirm attendance, prepare materials, and distribute posts evenly. It also collected independent variables for subject blocking, which ideally should have been completed beforehand for efficiency.
%
In \textit{introduction step}, the experiment begins. In this phase, we use \textit{guideline instruments} to present a few slides to the students, explaining the experiment's goal: identifying sustainability in the posts. We outline the procedure, provide a time estimate, and show them examples of how to find and read the posts.

%
In the \textit{experiment step}, students complete the \textbf{experiment questionnaire} as another \textit{measurement instrument} to identify sustainability in posts. Both groups receive similar questionnaires, but the flag group has links to sustainability flags, while the base group uses lecture slide sustainability definitions as the way of identifying sustainability. The \textit{wrap-up step} involves a final \textit{measurement instrument}, the \textbf{check questionnaire}. In this step, we ask a few questions about the students' overall experience with the tasks. While this data is not necessary for testing the hypotheses, it can help support the results. 
%
The questions assess the usefulness and understandability of the treatment (rated 1 = not at all to 5 = extremely) and the difficulty of identifying sustainability using the treatment (rated 1 = very easy to 5 = very hard) (see Table~\ref{tab:variables}). Students are also asked to describe their approach to identifying sustainability to understand their attitudes and methods. Finally, we present slides to wrap up, gather feedback, and reveal the treatments and their goals.
\subsubsection{Data collection procedure}
%
Data collection follows the planned, mentioned above procedure, primarily using \textit{online} questionnaires (explained above), and is completed within 105 minutes (a full course lecture split into two 45-minute sessions, skipping the 15-minute break). Grouping is done beforehand to save time. The introduction and briefing take 15 minutes, followed by 70 minutes for the experiment questionnaire, monitored by two experimenters to prevent communication and address questions. Afterward, data is automatically collected, followed by a 10-minute break and another 10 minutes for the check questionnaire and wrap-up slides.
\section{Flags Evaluation: Execution of the Experiment}
\label{sec:experiment-exec}
\begin{table}[tb]
\centering
\caption{Distribution of students' Background: Master Track, Year, and Experience across Base (B) Group and Flag (F) Group.}
\label{tab:blocking}
\fontsize{6}{7}\selectfont
\begin{tabularx}{\columnwidth}{|p{0.5cm} p{0.5cm} p{0.5cm}| p{0.5cm} p{0.5cm} p{0.5cm} |p{1cm} p{0.5cm} p{0.5cm}|}
\hline
\textbf{Track} & \textbf{B} & \textbf{F} & \textbf{Year} & \textbf{B} & \textbf{F} & \textbf{Experience } & \textbf{B} & \textbf{F} \\
\hline
X & 7 & 7 & 1st & 6 & 5 & Slides & 1 & 1 \\
Y & 1 & 1 & 2nd & 3 & 3 & Lecture & 7 & 6 \\ 
Z & 1 & 0 & & & & Prior & 1 & 1 \\
\hline
\end{tabularx}
\end{table}
\vspace{-2pt}
\subsection{Sample and preparation}
%
%
As described in \textit{Subjects} section~(\ref{sec:subjects}), the sample includes students from the same master's program but different tracks, all with knowledge of cloud computing, architecture, and sustainability. Data on these, along with their study year and participation, were collected via an anonymous demographic questionnaire, with random tags assigned to students. 

\subsection{Data collection performed and Validity procedure}
The data collection was carried out as planned in the study design, with no deviations. None of the participants dropped out of the experiment. It is worth noting that some participants finished 5-10 minutes earlier than the allocated time. However, we ensured that they did not disturb other participants.
With respect to validity, during the experiment, at least one researcher was always present to answer questions and ensure that the subjects did not communicate or use unauthorized materials (such as ChatGPT). Participants were carefully divided physically to prevent them from seeing the other group's treatment. All responses were collected automatically and online, ensuring there was no way for participants to view each other's opinions. All participants completed both the \textit{experiment} and the follow-up \textit{check} questionnaires.
\begin{table}[t]
    \centering
    \caption{Descriptive Statistics together with performance metrics for Base (B) and Flag (F) Groups}
    \label{tab:stats}
    \fontsize{6}{7}\selectfont
    \begin{tabular}{|p{0.35\columnwidth}|p{0.25\columnwidth}|p{0.25\columnwidth}|}
        \hline
        \textbf{Statistics/Group} & \textbf{B Group} & \textbf{F Group} \\ \hline
        \multicolumn{3}{c}{\textbf{Certainty Interval Question Statistics}}\\\hline
        \textit{Mode [Frequent Bin]} & [4.00, 4.50] &  [4.53, 5.00]\\
        \textit{Median} & 4 & 3.995 \\
        \textit{Standard Deviation} & 1.22 & 1.17 \\
        \textit{Mean} & 3.53 & 3.71 \\
        \textit{Variance} & 1.49 & 1.37 \\
        \textit{Min} & 0 & 0.27 \\
        \textit{Max} & 5 & 5 \\\hline
         \multicolumn{3}{c}{\textbf{Sustainability-relevant Question (Yes/No) Statistics}}\\\hline
        \textit{Frequency ("Yes)} & 152 & 129 \\
        \textit{Frequency ("No")} & 90 & 113 \\
        \textit{Mode} & Yes & Yes \\\hline
        \multicolumn{3}{c}{\textbf{performance Metrics against ground truth}}\\\hline
        \textit{Accuracy} & 0.52 & 0.71\\
        \textit{Recall} & 0.64 & 0.73\\
        \textit{Precision} & 0.53 & 0.72\\
        \textit{F1-Score} & 0.58 & 0.72\\\hline
    \end{tabular}
\end{table}
\section{Flags Evaluation: Analysis of the Experiment}
\label{sec:experiment-analysis}
\subsection{Descriptive statistics}
Following~\cite{wohlin2012experimentation}, for the certainty as interval data and sustainability relevance as nominal data, we report the suitable descriptive statistics in Table~\ref{tab:stats}. Moreover, Figures~\ref{fig:all-cerain} and~\ref{fig:all-yn} display the data distribution for the certainty interval and sustainability relevance (Yes/No). For the certainty interval (Fig.\ref{fig:all-cerain}), we use histogram bins with a Kernel Density Plot as recommended by~\cite{kitchenham2017robust}. In Fig.~\ref{fig:all-yn}, \textit{Yes/No} responses are reported in percentages rather than counts to account for the varied number of posts per student\cite{wohlin2012experimentation}.
\subsubsection{Certainty of identifying sustainability in the posts}
According to Table~\ref{tab:stats}, the F group exhibits a slight inclination towards higher certainty levels, as evidenced by a higher mean, a higher mode bin closer to the maximum value, and a marginally lower range and lower spread of intervals around the mean. While the differences between groups are subtle, F group appears to express slightly greater certainty, with ratings more concentrated around higher certainty levels.
\subsubsection{Frequency of identified sustainability-related posts}
When analyzing responses to sustainability relevance, a notable difference emerges between the groups. Base group members tended to classify posts as sustainability-relevant (``Yes'') more frequently, while F group ones were more inclined to mark posts as not relevant (``No''). This suggests that the F group may adopt a more conservative or selective perspective on sustainability relevance than the B group.
\vspace{-3pt}
\subsection{Dataset reduction}
\begin{figure}[t]
\centering
\includegraphics[width=0.6\columnwidth]{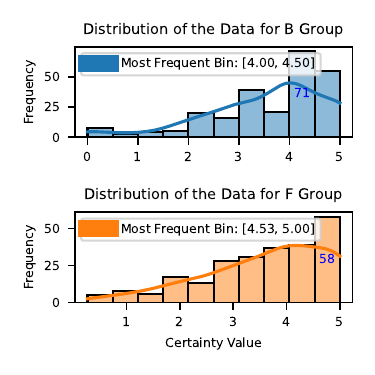}
\caption{The distribution of certainty intervals per Base (B) and Flag (F) group}
\label{fig:all-cerain}
\end{figure}
\begin{figure}[t]
\centering
\includegraphics[width=0.75\columnwidth]{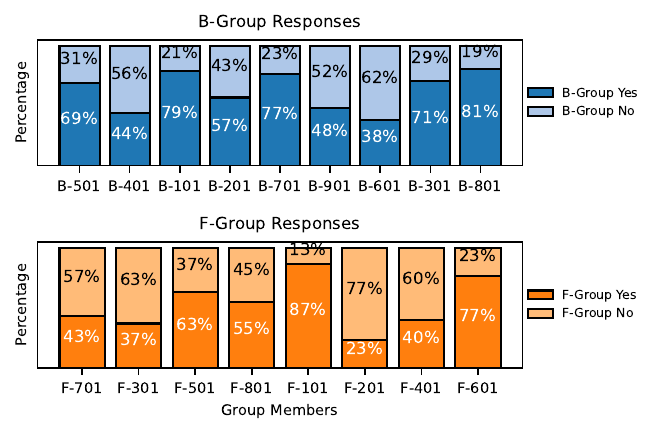}
\caption{The proportion of identified sustainability relevance per Base (B) and Flag (F) group}
\label{fig:all-yn}
\end{figure}
The certainty interval data shows no clear outliers, and both groups exhibit non-normal distributions as per Fig.~\ref{fig:all-cerain}. Fig.~\ref{fig:all-yn}, shows also no considerable outlier per group \textit{Yes/No}. Using these two figures and the given information in Table~\ref{tab:stats}, we decide not to remove any data point from the dataset.
\subsection{Hypothesis testing}
\subsubsection{Certainty of identifying sustainability in the posts}
The certainty data has these characteristics: sample size is not big, data is not normally distributed, data is interval, and the two groups of data are independent. Based on the recommendations by~\cite{wohlin2012experimentation}, the only appropriate statistical test for this not normally distributed data is the Mann-Whitney U test, a non-parametric method. The one-tailed Mann-Whitney U test on the certainty intervals for the B and F groups resulted in a U-value of 30,495 and a p-value of $\textbf{0.2145}$. Since the p-value is not below the standard threshold of 0.05, it does not provide strong evidence against the null hypothesis. This implies that the certainty levels of the flag group are not higher than those of the base group, or at least not statistically significant.
%
\subsubsection{The count (quantity) of identified sustainability-related posts}
For assessing the count of identified sustainability-related posts, we reference guidelines in~\cite{wohlin2012experimentation} and~\cite{nayak2011choose}, which suggest using Fisher's exact test. However, Fisher's test is appropriate only when each cell in the contingency table has values less than 30, which is not the case here. Therefore, we apply a one-tailed two-proportion z-test, an alternative for larger sample sizes~\cite{statstest}. The test yields a z-score of 2.118 and a p-value of $\textbf{0.0170}$. Since the p-value is below 0.05, we find significant evidence that the proportion of \textit{Yes} responses is higher in the B group than in the F group, indicating a greater tendency in the F group to not accept posts as relevant.
\subsubsection{Performance of both groups}
We calculated performance metrics for each group against the ground truth, as reported in Table~\ref{tab:stats}. According to these metrics, the F group consistently outperforms the B group. It demonstrates higher F1-score, accuracy, recall, and precision alignment with the ground truth results. To evaluate statistical power, we performed one-tailed z-tests on the proportions for accuracy, recall, precision, and F1-score, yielding z-scores of -4.201, -2.070, -4.092, and -3.206, with corresponding p-values of \textbf{0.000, 0.019, 0.000,} and \textbf{0.001}. As all p-values are below 0.05, we reject the null hypothesis ($H_{03}$) and conclude that the F group outperforms the B group across the mentioned measures.
\subsection{Ease of use of flags: insight from the check questionnaire}
\begin{figure*}[t]
\centering
\includegraphics[width=1.75\columnwidth]{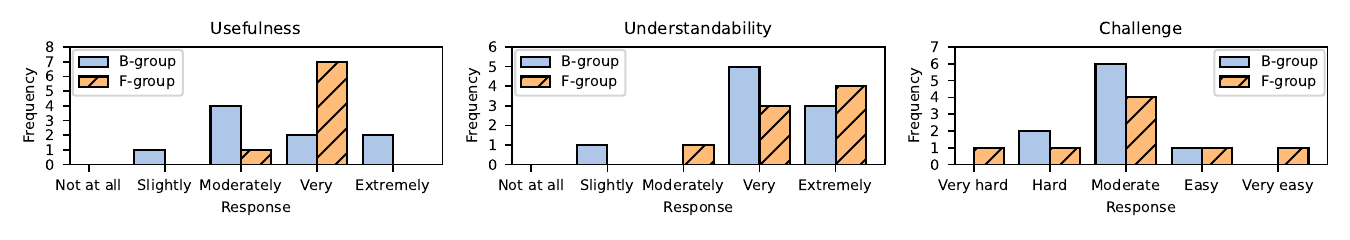}
\caption{The responses of both groups' subjects for the questions about the ease of adoption of their provided treatment. Groups: Base (B) and Flag (F).}
\label{fig:all-check}
\end{figure*}

We refer to the results from the check questionnaire presented in Fig.~\ref{fig:all-check} to learn about ease of use. For ordinal data, central tendency is typically measured using the mode or median~\cite{MARSHALL2010e1}. However, calculating the median is unsuitable here due to the F group's even size (8). In this case, we have to take the arithmetic mean but the ordinal scale can not support arithmetic operations like mean~\cite{Scribbr} and we are only left with mode. This figure, which also shows mode per group illustrates that, while both groups (F and B) perceive the task of identifying sustainability as comparably challenging, participants in the F group view sustainability flags as more useful and slightly more understandable for this purpose than those in the B group. This comes from the fact that the mode of usefulness chart for the F group is \textit{very useful} while for the B group, this is \textit{moderately useful}. The situation on understandability for the F group is only slightly better than the B group. This suggests that the F group may find greater clarity and usefulness in using sustainability flags for identification, as compared to the B group.

\section{Discussion}
\label{sec:discussion}
\subsection{Interpretation of the Experiment}
The $H_{01}$ and $H_{1}$ hypotheses address the certainty level in identifying sustainability. The results do not provide sufficient evidence to confirm or reject the null hypothesis $H_{01}$, meaning we cannot show that using sustainability flags significantly impacts subjects' certainty in identifying sustainability. This finding was unexpected, as we anticipated the flag group would report significantly higher certainty due to relying on defined flags rather than broad definitions. These findings may however lack statistical power due to the relatively small sample size. On the other hand, the flag group displayed lower variance, suggesting greater consistency in the certainty of their replies. This consistency is further reflected in the smoother, more uniform distribution curve of the flag group (Fig.~\ref{fig:all-cerain}). Additionally, the mode indicates a slightly higher certainty level in the flag group than in the base group.

The $H_{02}$ and $H_{2}$ hypotheses concern the number of posts identified as sustainability-related. Here, we successfully reject the null hypothesis, showing that the flag group accepts fewer posts than the base group. This is notable because, as expected, the flag group uses defined flags to identify sustainability, while the base group relies on personal heuristics and broad sustainability definitions, which are less restrictive. This finding gains further meaning when we consider the performance metrics (from $H_{03}$ which is successfully rejected), which indicate better accuracy and precision for the flag group.  \textcolor[HTML]{0066ff}{These results suggest that the flag group is more rigorous in identifying sustainability-related posts while maintaining higher accuracy, likely due to the structured guidance provided by the flags.}

The last point \cmmnt{$H_{4}$ hypothesis} concerns the ease of adopting sustainability flags for identifying sustainability. The reason the flag group may find this treatment more useful than the base group is likely due to the specific guidance provided: the flags explicitly outline the points and concepts to identify in the text, whereas the base treatment offers only broad definitions, leaving much to individual interpretation. Although the flags are clearer than general sustainability definitions, applying them to assess sustainability remains challenging. Some may question the benefit if both treatments are still difficult to use.  \textcolor[HTML]{0066ff}{Here, the intended advantage of the flags is not only to simplify the task but to improve the feasibility and accuracy of identifying sustainability in posts where the term ``sustainability'' is rarely or never mentioned.} This response leads us to an additional interesting point that we aim to explore in the following.

At the end of the check questionnaire, we asked both groups: ``Can you briefly explain how you approached identifying sustainability in the posts?'' We analyzed their responses qualitatively. We found that all members of the flag group focused exclusively on identifying sustainability by searching for specific terms provided in the treatment. In contrast, subjects in the base group used a variety of methods, such as focusing on a particular sustainability dimension, identifying the main idea of the post, and looking for indicators like long-term benefits, financial aspects, code efficiency, and other concepts. In contrast, each member in the base group relied on their own reasoning and inferred or felt sustainability-related aspects based on personal heuristics, which are not necessarily replicable by others or by a computer program. \textcolor[HTML]{0066ff}{The takeaway here is that the introduction of sustainability flags seems to streamline the classification of posts for sustainability, at least by human raters. Searching for specific concepts and terms associated with them (i.e., flags) is far more feasible and consistent than searching for abstract concepts or creating subjective heuristics. This approach enhances replicability and could facilitate the automation of sustainability classification and analysis in posts.}
\vspace{-0.25pt}
\subsection{Implications for Researchers and Practitioners}
For \textbf{researchers}, our study represents an initial step toward a method for sustainability identification, particularly for those exploring sustainability from software repositories like Q\&A, specifically about architecting on cloud environments. We encourage \textbf{practitioners} to consider the context and nature of the sustainability flags derived from existing BPs. These practices, widely overlapping with performance efficiency strategies familiar to most software practitioners, should be promoted and adopted. Sustainability flags are particularly accessible and intuitive as they align closely with well-understood concepts of performance efficiency. Moreover, as these flags and BPs demonstrate, there is significant overlap between sustainability and performance efficiency, and possibly with other architectural and design concerns yet to be explored. We hope this sparks awareness among practitioners, encouraging them to reflect on the sustainability impact of their decisions, especially in the realm of performance efficiency. Overall, our study conveys an essential message to practitioners:  \textcolor[HTML]{0066ff}{sustainability is probably already part of your decision-making discussions, even if it is not recognized as such. Using flags can start giving you awareness}.
\subsection{Lessons Learned and Insights from the Study}
When reviewing the extracted sustainability flags, we notice significant overlap--especially in ``resource efficiency,'' ``network efficiency,'' and ``storage efficiency''--with performance efficiency tactics like those in~\cite{bass2021}, such as managing work requests, reducing computational overhead, improving resource usage efficiency, and etc. This overlap likely occurs because many sustainability BPs align with established performance efficiency strategies. This insight conveys an important message: \textcolor[HTML]{0066ff}{promoting sustainability does not always require new approaches but rather a shift in focus. Recognizing that key aspects like\textit{ performance efficiency}, already fundamental to software design, also contribute to sustainability can be impactful.} Instead of creating new tactics/BPs, we may first raise awareness of the dual benefits existing solutions offer, an approach we plan to explore further.

While sustainability flags show promise in identifying sustainability, they require improvement. As shown in Fig.~\ref{fig:design-sci}, addressing a knowledge problem often reveals a new design problem. In our case, the knowledge problem (experiment) shows improved \textit{certainty}, but not significantly. This highlights the need to refine/augment the flags to guide labelers in identifying sustainability more certain.  \textcolor[HTML]{0066ff}{As future work, our first step is to develop a concrete codebook for each flag to support this process,} staring with steps mentioned in~\cite{soliman2024exploring}.

The current arrangement of flags also does not explicitly associate them with specific sustainability dimensions, as their origin--best practices--lacks such connections. While BPs show only weak associations with the environmental dimension, we believe they inherently relate to multiple sustainability dimensions, which are not reflected in the context provided by cloud providers. Therefore, we do not limit the flags to the environmental dimension. We also do not assign them to multiple dimensions at this stage. This is because this assignment requires expert input through case studies or interview running, which fall outside the scope of this study. \textcolor[HTML]{0066ff}{Exploring these associations will be a key focus of future work.}
\section{Threats to the Validity}
\label{sec:ttv}
For reproducibility purposes, the comprehensive~\href{https://figshare.com/s/90b8b9a72560d5a23086}{\textbf{replication package}}\footnote{https://figshare.com/s/90b8b9a72560d5a23086} of this research is made available online. In the following, we discuss potential risks, mitigation measures, and limitations of our work along the lines of the two methodological instruments we used for answers RQ1 and~2.
To address the limitations of this study, we consider the classification presented by~\cite{wohlin2012experimentation}, which is discussed based on Cook and Campbell~\cite{cook1979}.
\subsubsection{Internal Validity} 
\label{sec:ttvInternal}
If a relationship is observed between the treatment and outcome, we must ensure it is causal and not influenced by uncontrolled or unmeasured factors. The questionnaire is identical for both groups, with only two simple questions per post. However, the dataset may introduce two threats: first, the number of posts per subject varies slightly (28–31). To address this, we balanced the posts per subject by creation year and length, ensuring each subject sees posts from nearly every year and evenly distributes longer posts (over 1000 words). Second, if the dataset had an excess of flag-related posts, it could bias results, though this is unlikely, as the classification of the posts in this dataset was independent of both sustainability flags and sustainability definitions. Additionally, since all posts are from the Software Engineering forum on StackExchange, a context-specific influence may still exist, which we cannot fully mitigate.

Human performance variability could influence the experiment, yet this was mitigated by effective blocking and creating two balanced groups. The base group had no chance to mimic or learn from the flag group, as they were seated far apart in a large hall, preventing any visual access to each other's materials. Both groups received initial guidance, and while one might suggest that simpler instructions could sway performance, we believe the complexity of both guides was well-balanced. The flag group's guide was more detailed with examples, whereas the base group's guide, though shorter and using visuals, was inherently broader due to the general nature of sustainability definitions.

To create reliable labels for the \textbf{ground truth dataset}, we had four researchers independently review and label most of it, specifically, 97 out of the total number of 121 posts (80\%). This process ensured that each of these posts received two initial labels.
To check how well the researchers agreed with each other, we calculated Cohen's Kappa. The resulting value was \textbf{0.6743}, which indicates a substantial level of agreement.

Following the initial labeling, the researchers discussed any disagreements they had. The goal of these discussions, held in multiple consensus meetings, was to arrive at a shared understanding of what makes a post related to sustainability. Once this common understanding was reached, one of the researchers labeled the remaining posts. the ground truth is not available to the students as it is produced beforehand.

Another internal validity might have happened during conducting \textit{thematic analysis}. To minimize biases in code and theme extraction, we rigorously follow the thematic analysis approach~\cite{cruzes2011}, involving three researchers who collaboratively review emerging themes until full consensus is reached. Based on the methodology, the disagreements have to be resolved either by consensus or arbitration by an additional independent researcher. In our case, always while two researchers are involved, the full consensus has been made and we do not need a third researcher. Moreover, we make sure that in every step that more than one researcher is involved, the \textbf{full agreement} has been met. At the end, a protocol, based on~\cite{cruzes2011} and~\cite{braun2012thematic}, shared with all researchers beforehand, ensures consistent adherence to the method throughout the process. The protocol is also shared in the replication package.
%
%
\subsubsection{Construct Validity} This validity is about the suitability of the study design for the theory behind the experiment. We make sure what \textit{effectiveness} and the \textit{ease of use} mean in the experiment by means of identifying clear dependent variables and how we measure them. Thus, we believe the pre-operational explication of the construct is adequate, and in this way we also avoid mono-method bias.
In this study, the dataset could however introduce mono-operation bias. However, we believe this risk is minimized for two reasons: first, inspection occurs at the post level, with each subject viewing multiple unique posts; second, while drawn from a single forum, the dataset spans from 2016 to 2024, covering diverse perspectives on cloud architecture. We assume this dataset reasonably represents cloud architecture discussions, particularly for analysis, synthesis, and implementation phases~\cite{ahmadisakha2024architecting}. Nonetheless, we acknowledge that this threat may not be fully mitigated.
%
%
\subsubsection{External Validity} This validity focuses on the generalizability of the results to a broader population. Although students are not ideal subjects, they are considered future software professionals and thus relatively close to the target population of interest of non-experts on sustainability, as noted in~\cite{kitchenham2002preliminary}. Choosing the students as subjects is not considered a threat if we can carefully group and choose their characteristics for this experiment~\cite{ralph2020empirical}, something that we carefully perform by means of the rigorous selection of the course and the students and spreading them between the groups; thus we think we could show the appropriateness of students to perform this experiment. The experimental subjects may nevertheless not fully represent the larger population; in this case, students are used as a worst-case scenario. We assume that if the treatment is effective for them, it may be even more effective for experienced software architects. Our subjects have limited cloud architecting knowledge, which restricts the experiment's generalizability to cloud-based architecture contexts though.

Another potential threat is that the dataset and problem analyzed may be too simple or unrealistic for broader generalization. However, the dataset consists of real posts from an active Q\&A forum frequently used by software architects as a primary resource for architectural answers~\cite{de2022developers}, ensuring its relevance and realism. 
\subsubsection{Conclusion Validity}
\label{sec:ttv-conclusion_val}
This validity is concerned with the relationship between the treatment and the outcome. A key threat is the lack of statistical significance for the certainty-related findings (the first hypothesis), which raises the risk of drawing an erroneous conclusion for this hypothesis. However, additional evidence from the study, including the second and third hypotheses with high statistical significance, supports the efficacy of using sustainability flags. \cmmnt{For the third and fourth hypotheses, where no specific null hypothesis is defined, traditional statistical power analysis may be less applicable. Instead, focusing on effect sizes should suffice, particularly given the exploratory nature of these hypotheses~\cite{gelman2013garden}.} While there may be concerns about the assumptions underlying the chosen statistical tests, we have carefully reviewed and plotted the data, and consulted a statistical expert. In the end, someone might argue that the measures used are not so reliable since the answers to the questions for each post may be subjective. However, we obtained two independent opinions per post in each of the F and B groups, providing multiple perspectives without any mutual influence.
\vspace{-5pt}
\section{Related Works}
\label{sec:related}
Software engineering plays a crucial role to create sustainable software systems. To put it into action, researchers tried to (re)define the concept for software systems, as per~\cite{penzenstadler2012sustainability, calero2013sustainability} and frame it with some principles and notes in the Karlskrona manifesto~\cite{becker2014karlskrona}. The filed then tried to bring solutions first with the concept of green software~\cite{kern2015impacts, hindle2016green} and expanded to incorporate additional dimensions~\cite{garcia2018interactions, wolfram2017sustainability}. Since software architecture is the foundation of any software system, and the most critical decisions are made at this stage~\cite{jansen2005software}, the focus on sustainability in software architecture also gained importance, particularly after sustainability was recognized as a software quality in 2015~\cite{lago-2015}. %
Since 2015, studies have shown a focus on sustainability in software architecture. For instance, \cite{betz2015sustainability} examines sustainability debt in software engineering, while \cite{venters-2018, venters2023sustainable} reviews sustainability in software architecture with a technical focus. \cite{lago-2019} presents architectural maps to incorporate sustainability dimensions, alongside a quality model developed in \cite{condori2018characterizing, condori2020action}. In cloud architecture, \cite{procaccianti-2015} and \cite{vos2022architectural} suggest tactics for environmental sustainability, while \cite{farshidi2018provider, farsh2018service} evaluate cloud providers and services to support sustainable choices.

The field of content analysis for sustainability is sparse, especially beyond environmental sustainability, with existing work largely focused on energy consumption. This focus typically involves term-based searches to identify energy-related posts, followed by manual analysis \cite{malik2015going}, \cite{jin2024practitioners}, \cite{farooq2019bigdata}, \cite{pinto2014mining}, \cite{burac2024decoding}, \cite{albonico2021mining}. While term-based classification works well for energy-related topics due to an established, widely accepted terminology, it is not as feasible across other sustainability dimensions. The only work addressing sustainability in Q\&A forums is \cite{ahmadisakha2024mining}, which uses a term-based approach to identify sustainability posts. However, it lacks clarity on effectiveness and performance and uses quality requirements as a proxy to classify posts rather than sustainability dimensions definition.

\section{Conclusion and Future Work}
\label{sec:conclusion}
Our study aims to identify sustainability discussions in Q\&A forums. Using only sustainability definitions for classification is ineffective, so we derive the concept of sustainability flags as pointers to relevant posts based on the widely adopted hyperscalers' best practices on the topic of sustainability. We hypothesize that these flags assist the identification task, especially for non-experts on the topic. In a controlled experiment with students, results show the flag-using group outperforms the control group in accuracy, precision, recall, and F1-score, with a higher rejection rate supporting the flags' effectiveness. The flag group finds the task as challenging as the base group, but rates the flags as more useful and slightly more understandable than only definitions. The next step in our future work would be augmenting the flags by developing a codebook for each of them and further exploring and defining each flag's association with the four sustainability dimensions.
%
\bibliographystyle{ieeetr}
\bibliography{REF}
\end{document}